\documentclass[10pt,twocolumn,letterpaper]{article}

\usepackage{cvpr}
\usepackage{times}
\usepackage{epsfig}
\usepackage{graphicx}
\usepackage{amsmath}
\usepackage{amssymb}
\usepackage{mathtools}
\usepackage{xfrac}
\usepackage{subcaption}

\usepackage[breaklinks=true,bookmarks=false]{hyperref}
\cvprfinalcopy 

\ifcvprfinal\fi

\newcommand{\bias}{\operatorname{B}}
\newcommand{\gain}{\operatorname{G}}
\newcommand{\curve}{\operatorname{C}}

\begin{document}

\title{A Convenient Generalization of Schlick's Bias and Gain Functions}

\author{Jonathan T. Barron\\
{\tt\small barron@google.com}
}

\maketitle

\begin{abstract}
   We present a generalization of Schlick's bias and gain functions --- simple parametric curve-shaped functions for inputs in $[0, 1]$. Our single function includes both bias and gain as special cases, and is able to describe other smooth and monotonic curves with variable degrees of asymmetry.
\end{abstract}

Schlick's bias and gain functions~\cite{schlick1994fast} are simple tools for modeling smooth and monotonic curve-shaped mappings from $[0,1]$ to $[0, 1]$. These functions were presented as inexpensive rational alternatives to the similarly-shaped bias and gain functions presented by Perlin~\cite{perlin1989hypertexture}. Though bias and gain were originally developed for use in procedural texture generation they are often used as general purpose ``easing'' functions, or as a means to manipulate interpolation weights.

\begin{figure*}[t!]
    \centering
    \subcaptionbox{$y = \bias(x, a$)\label{subfig:bias}}{
    \includegraphics[trim=0.3in 0.3in 0.3in 0.7in, clip, width=2.5in]{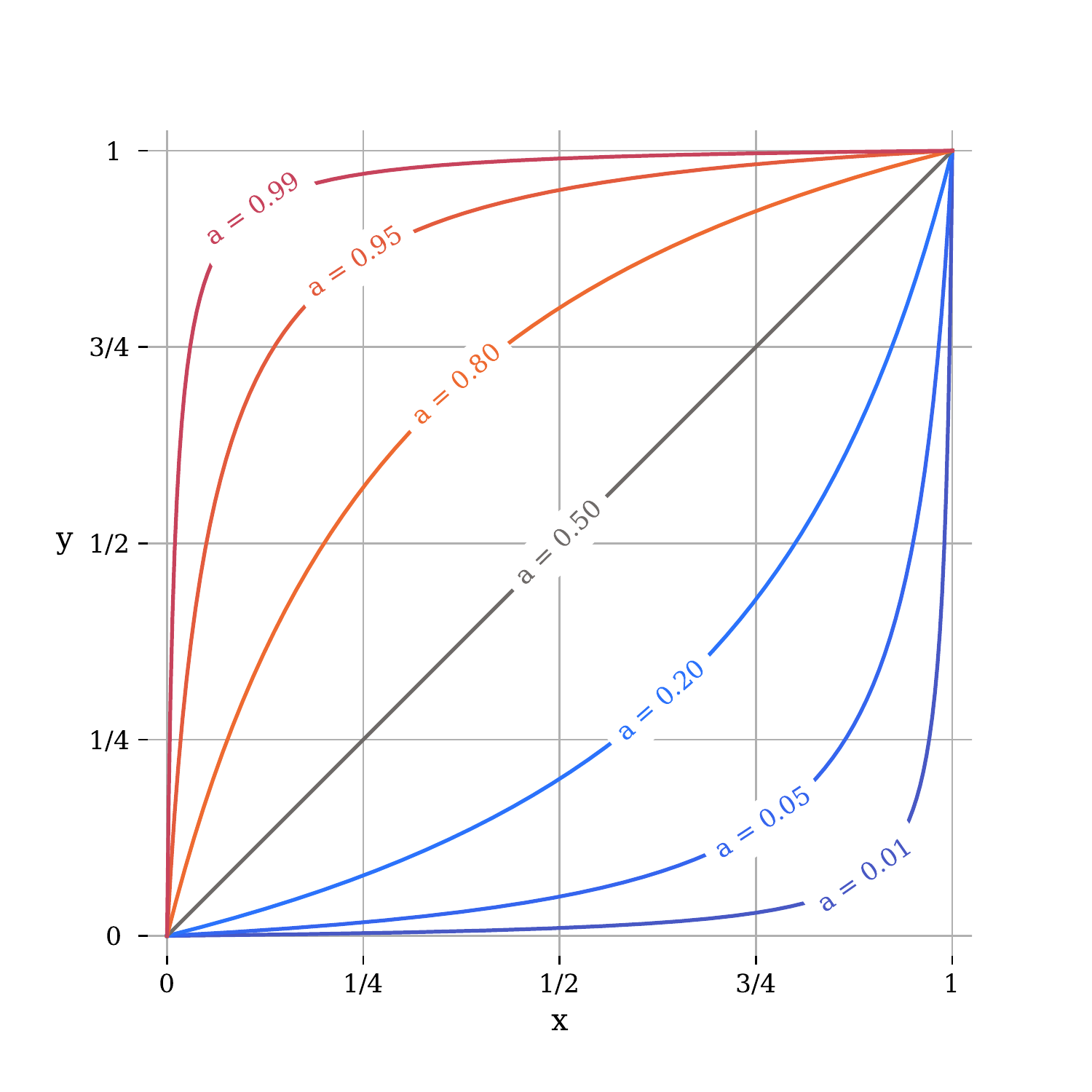}}
    \subcaptionbox{$y = \gain(x, a$)\label{subfig:gain}}{
    \includegraphics[trim=0.3in 0.3in 0.3in 0.7in, clip, width=2.5in]{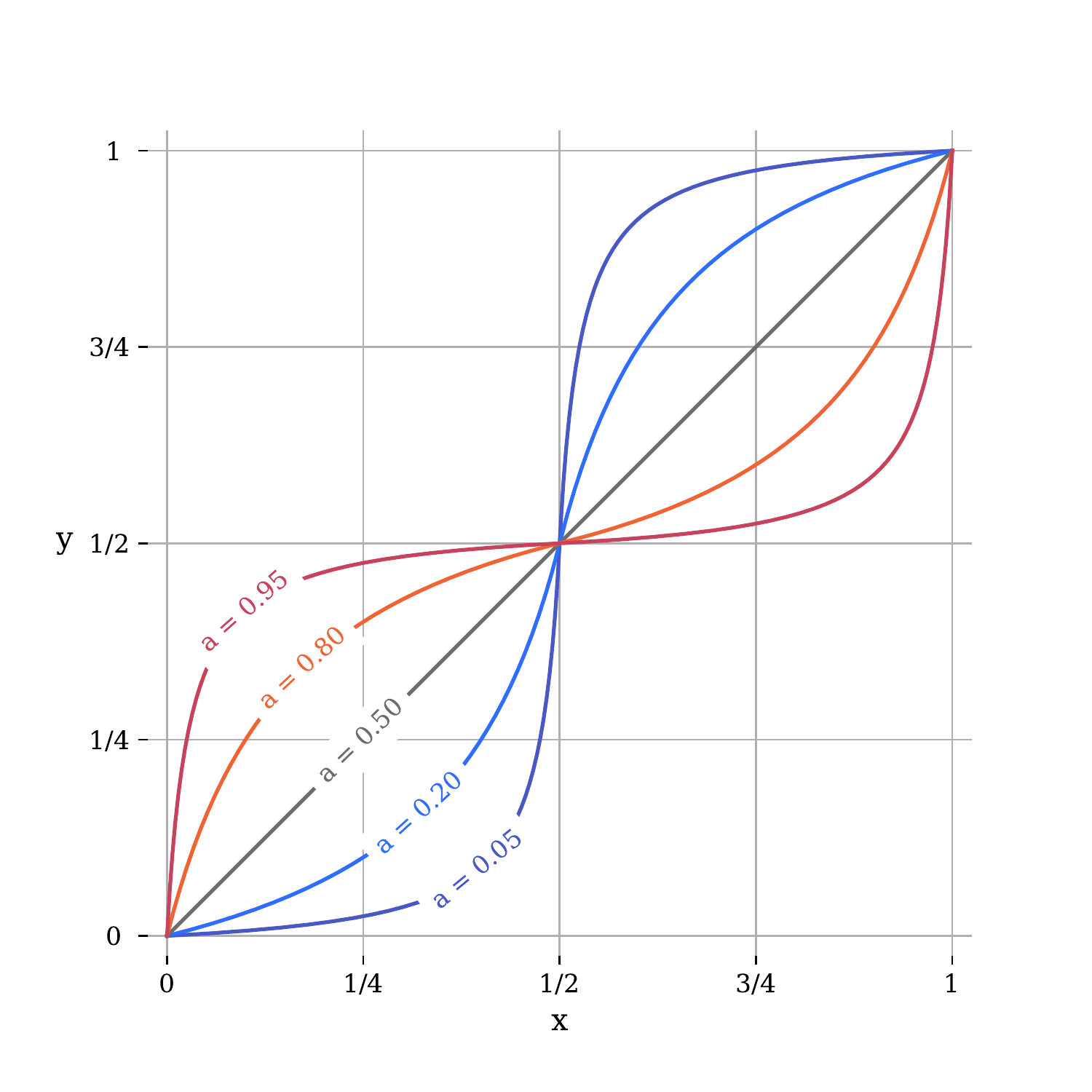}}
    \caption{Visualizations of Schlick's bias and gain functions for different values of $a$.}
    \label{fig:bias}
\end{figure*}

Schlick's bias and gain are respectively defined as:
\begin{align}
    \bias(x, a) &= \frac{x}{ (1/a - 2)(1-x) + 1}\,, \\
    \gain(x, a) &= \begin{dcases}
        \frac{\bias(2x, a)}{2}        & \mbox{if } x < \sfrac{1}{2} \\
        \frac{\bias(2x-1,1-a)+1}{2} & \mbox{if } x \geq \sfrac{1}{2}\,,
        \end{dcases}
\end{align}
where the input $x$ is in $[0, 1]$, the shape of each function is determined by one parameter $a \in (0, 1)$, and the output is guaranteed to be in $[0, 1]$.
See Figure~\ref{fig:bias} for visualizations of bias and gain for different values of $a$.
The bias function $\bias$ ``biases'' the input towards either $0$ or $1$, while the gain function $\gain$ pushes or pulls the input towards or away from $\sfrac{1}{2}$.
The bias function is a single curve that is symmetric across $y = 1-x$; setting $a<\sfrac{1}{2}$ biases the output towards $0$, setting $a=\sfrac{1}{2}$ yields the identity function, and setting $a>\sfrac{1}{2}$ biases the output towards $1$.
The gain function is two scaled and shifted bias curves joined at $x=\sfrac{1}{2}$, where each curve's shape parameter is one minus that of the other. 

Our generalization of bias and gain is as follows:
\begin{equation}
\curve(x, s, t) = \begin{dcases}
        \frac{tx}{x+s(t-x)+\epsilon}        & \mbox{if } x < t \\
        \frac{(1-t)(x-1)}{1-x-s(t-x)+\epsilon} + 1 & \mbox{if } x \geq t\,,
        \end{dcases}
\end{equation}
where the input $x$ is in $[0, 1]$ and the shape of the function is determined by two parameters $s \geq 0$ and $t \in [0, 1]$ ($\epsilon$ is machine epsilon, and just prevents division by zero when $t$ is $0$ or $1$). The ``threshold'' parameter $t$ controls the value of $x$ where the curve reverses it shape, which is also the only input location (besides $x=0$ and $x=1$) where the output of the curve is guaranteed to be equal to its input.
The ``slope'' parameter $s$ controls the slope of the curve at the threshold $t$, and is exactly the derivative of $\curve(x, s, t)$ with respect to $x$ at $x=t$.
This parameterization allows a practitioner to shape the curve similarly to how one might adjust a Hermite spline.

We can reproduce Schlick's bias two ways, one in which $t=0$ and the slope is small (or large) and another in which $t=1$ and the slope is large (or small).
\begin{equation}
\bias\left(x, \sfrac{1}{(s+1)}\right) = \curve\left(x, \sfrac{1}{s}, 0\right) = \curve(x, s, 1)\,.
\end{equation}
We can also reproduce Schlick's gain by setting $t=\sfrac{1}{2}$.
\begin{equation}
\gain\left(x, \sfrac{1}{(s+1)}\right) = \curve(x, s, \sfrac{1}{2})\,.
\end{equation}
Our generalization  exhibits similarly symmetries as Schlick's bias and gain functions:
\begin{equation}
    \curve(x, s, t) =  1 - \curve(1-x, s, 1-t)\,.
\end{equation}
We can invert our function by setting $s$ to its reciprocal:
\begin{equation}
 x = \curve\!\big(\!\curve(x, s, t), \sfrac{1}{s}, t\big)\,.
\end{equation}

\begin{figure*}[b!]
    \centering
    \includegraphics[trim=1.5in 1.5in 0.9in 1.5in, clip, width=5.5in]{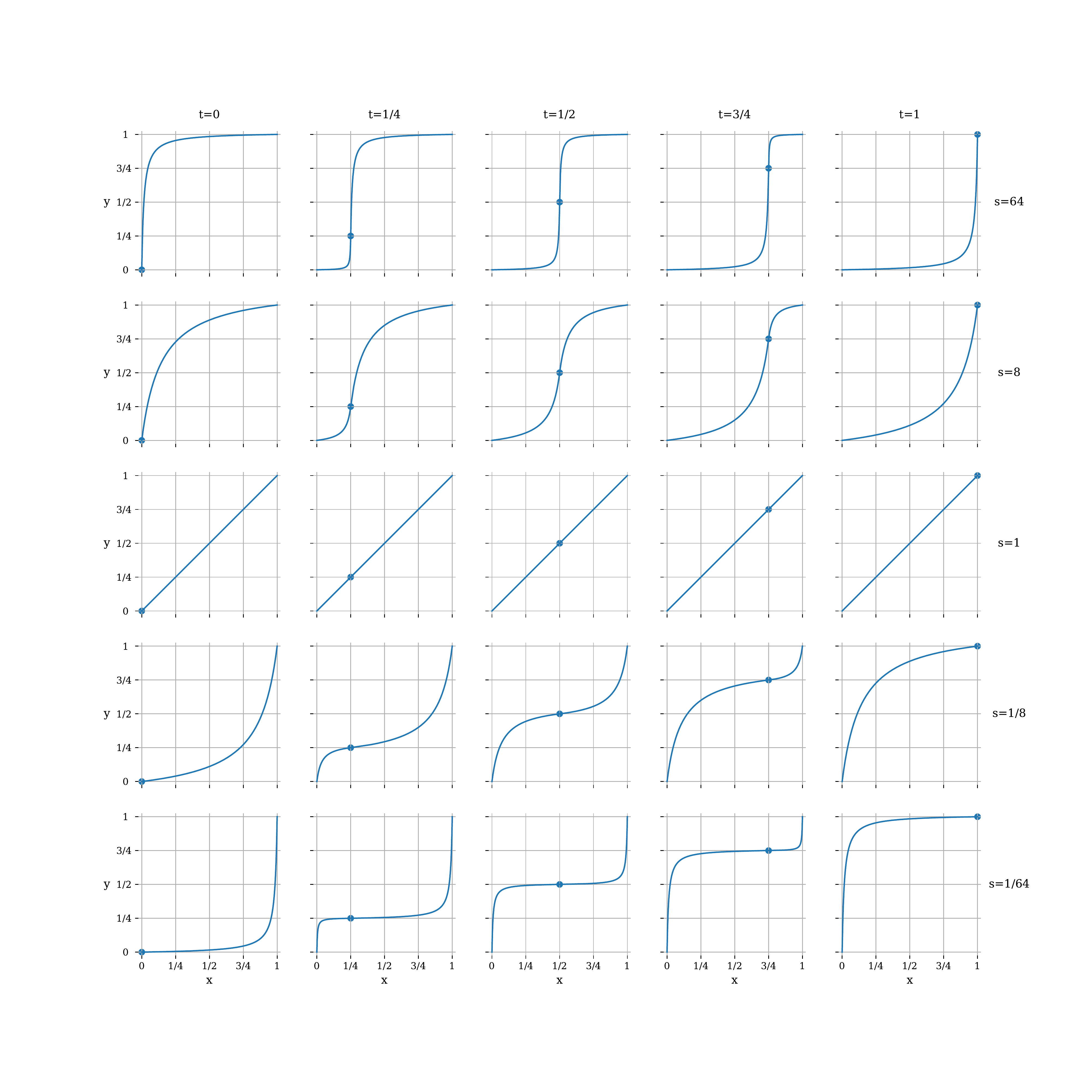}
    \caption{A ``small multiples'' visualization of $y = \curve(x, s, t)$ for different values of $s$ (rows) and of $t$ (columns). A knot indicates where $x=t$. When $t$ is $0$ or $1$ (first column, last column) our generalization reproduces Schlick's bias, and when $t$ is $\sfrac{1}{2}$ (middle column) it reproduces Schlick's gain.}
    \label{fig:curves}
\end{figure*}

Unlike Schlick's bias and gain functions, our generalization is able to describe asymmetric gain-like curves by setting $t$ to values other than $\sfrac{1}{2}$, as shown in Figure~\ref{fig:curves}.

{\small
\bibliographystyle{ieee_fullname}
\bibliography{curve}
}

\end{document}